\documentclass[referee,pdflatex,sn-vancouver,Numbered]{sn-jnl}
\usepackage{natbib}

\usepackage{graphicx}%
\usepackage{multirow}%
\usepackage{amsmath,amssymb,amsfonts}%
\usepackage{amsthm}%
\usepackage{mathrsfs}%
\usepackage[title]{appendix}%
\usepackage{xcolor}%
\usepackage{textcomp}%
\usepackage{manyfoot}%
\usepackage{algorithm}%
\usepackage{natbib}
\usepackage{algorithmicx}%
\usepackage{algpseudocode}%
\usepackage{listings}%
\usepackage{enumerate}
\usepackage{caption}
\usepackage{makecell}
\usepackage{booktabs}
\usepackage{comment}

\theoremstyle{thmstyleone}%
\theoremstyle{thmstyletwo}%

\theoremstyle{thmstylethree}%

\begin{document}

\title[FDA of postprandial CGM in AEGIS]{ Beyond Scalar Metrics: Functional Data Analysis of Postprandial Continuous Glucose Monitoring in the AEGIS Study}

\author*[1]{\fnm{Marcos} \sur{Matabuena}}\email{mmatabuena@hsph.harvard.edu}

\author[2]{\fnm{Joseph} \sur{Sartini}}\email{jsartin1@jhu.edu}
\equalcont{MM and JS share the first author position.}

\author[3]{\fnm{Francisco} \sur{Gude}}\email{francisco.gude.sampedro@sergas.es}

\affil*[1]{\orgdiv{Biostatistics Dept.}, \orgname{Harvard University}, \orgaddress{\street{677 Huntington Ave}, \city{Boston}, \postcode{02115}, \state{MA}, \country{United States}}}

\affil[2]{\orgdiv{Biostatistics Dept.}, \orgname{Johns Hopkins University}, \orgaddress{\street{615 N Wolfe St}, \city{Baltimore}, \postcode{21205}, \state{MD}, \country{United States}}}

\affil[3]{\orgdiv{Dept. of Medicine}, \orgname{Universidad de Santiago de Compostela}, \orgaddress{\street{Praza do Obradoiro}, \city{Santiago de Compostela}, \postcode{15705}, \country{Spain}}}




\abstract{ 
Postprandial glucose collected through continuous glucose monitoring (CGM) provides critical information for assessing metabolic capacity and guiding dietary recommendations. Traditional approaches summarize these data into scalar measures, such as 2-hour AUC or peak glucose, potentially overlooking temporal dynamics. We propose analyzing entire CGM trajectories using multilevel functional data analysis (FDA), which accounts for the smooth, hierarchical nature of glucose responses. Applying these methods to AEGIS study participants without diabetes, we illustrate how FDA characterizes variability in postprandial responses and links dietary/patient characteristics to glucose dynamics. We further extend the R² metric to hierarchical functional models to quantify explanatory power. Our results show that dietary effects vary across the 6-hour postprandial window—for example, fiber blunts responses after ~90 minutes, while fats reduce early rises within 50 minutes. Moreover, metabolic responses differ between normoglycemic and prediabetic individuals. These findings demonstrate that functional approaches reveal temporal and stratified insights into postprandial glucose regulation that scalar methods cannot capture.}

\keywords{Continuous glucose monitoring, Postprandial glucose, Functional data analysis, Glucose metabolism, Hierarchical modeling}



\maketitle

\section{Background}\label{intro}

While there is substantial literature regarding the impact of meal composition on scalar summaries of postprandial glucose response (PPGR), few studies have examined how the continuous glucose time series during the whole postprandial window is impacted by diet \cite{mozaffarian_dietary_2008, berry_human_2020, merino_validity_2022, jagannathan_oral_2020}. The A Estrada Glycation and Inflammation Study (AEGIS) provides an opportunity to study this relationship in a free-living setting through concurrent collection of meal timing and continuous glucose monitoring (CGM) data. Recent advances in wearable technology and smartphones have revolutionized the collection of physiological time series data in large samples \cite{kellogg_personal_2018, dunn_wearables_2018, johnson_wearable_2023}, and continuous glucose monitoring technology is no different \cite{rodbard_continuous_2016-1, ebekozien_technology_nodate,battelino_clinical_2019, matabuena_glucodensities_2021, matabuena_kernel_2022, beck_validation_2018, rodbard_glucose_2018, cui_investigating_nodate}. CGM provides real-time glycemic response data helpful for diabetes management, screening, and overall assessment of glucose metabolism \cite{ben-yacov_gut_2023, hall_glucotypes_2018, matabuena_kernel_2022}. To our knowledge, only a few studies have attempted to holistically characterize time-dependent metabolic response patterns in the high-resolution CGM time series collected during the postprandial period, and many such studies focus just on a subset of features such as peaks \cite{brand-miller_glycemic_2009}. Many prior nutritional studies focus primarily on metabolites \citep{wang_predicting_2023}, collect serum blood glucose a fixed time after meals \citep{song_individual_2023}, or summarize CGM data into simple scalar summaries such as 2 hour AUC, 2 hour peak glucose, or 2 hour mean glucose \citep{zeevi_personalized_2015, jarvis_continuous_2023, trouwborst_cardiometabolic_2023}. Analysis of the entire postprandial glucose response trajectories ensures that no clinically information is lost during aggregation.

CGM devices facilitate characterization of how postprandial glucose trajectories vary by diet and individual. Care must be taken to appropriately model the corresponding glucose data. Differential equations (DE) frameworks for modeling these data based upon physiological principals have been proposed, but many instances require also observing the related hormones of glucagon and insulin \cite{urbina_mathematical_2020, maas_physiology-based_2015, shi_modeling_2020,  trajanoski_fuzzy_1996, holtschlag_state-space_1998}. Alternatively, one can leverage the rich literature of DE-based minimal models governing just glucose dynamics \citep{eichenlaub_minimal_2019, ng_parsimonious_2022, de_gaetano_modeling_2021}. Although physiologically well--motivated, these models can be computationally demanding and often provide no way to estimate time--dependent statistical associations between predictors and the glucose response. These issues are compounded when the data have a hierarchical structure (e.g., multiple postprandial responses per individual).

In healthy populations, analyzing postprandial CGM responses is a valuable way to assess an individual’s metabolic capacity, and several machine-learning studies have recently addressed this topic \cite{metwally_prediction_2024,montaser_essential_2022,montaser_predicting_2023}. However, these studies primarily focus on combining pools of scalar CGM and non--CGM biomarkers.

\begin{figure}[ht!]
  \centering
  \includegraphics[width=10cm]{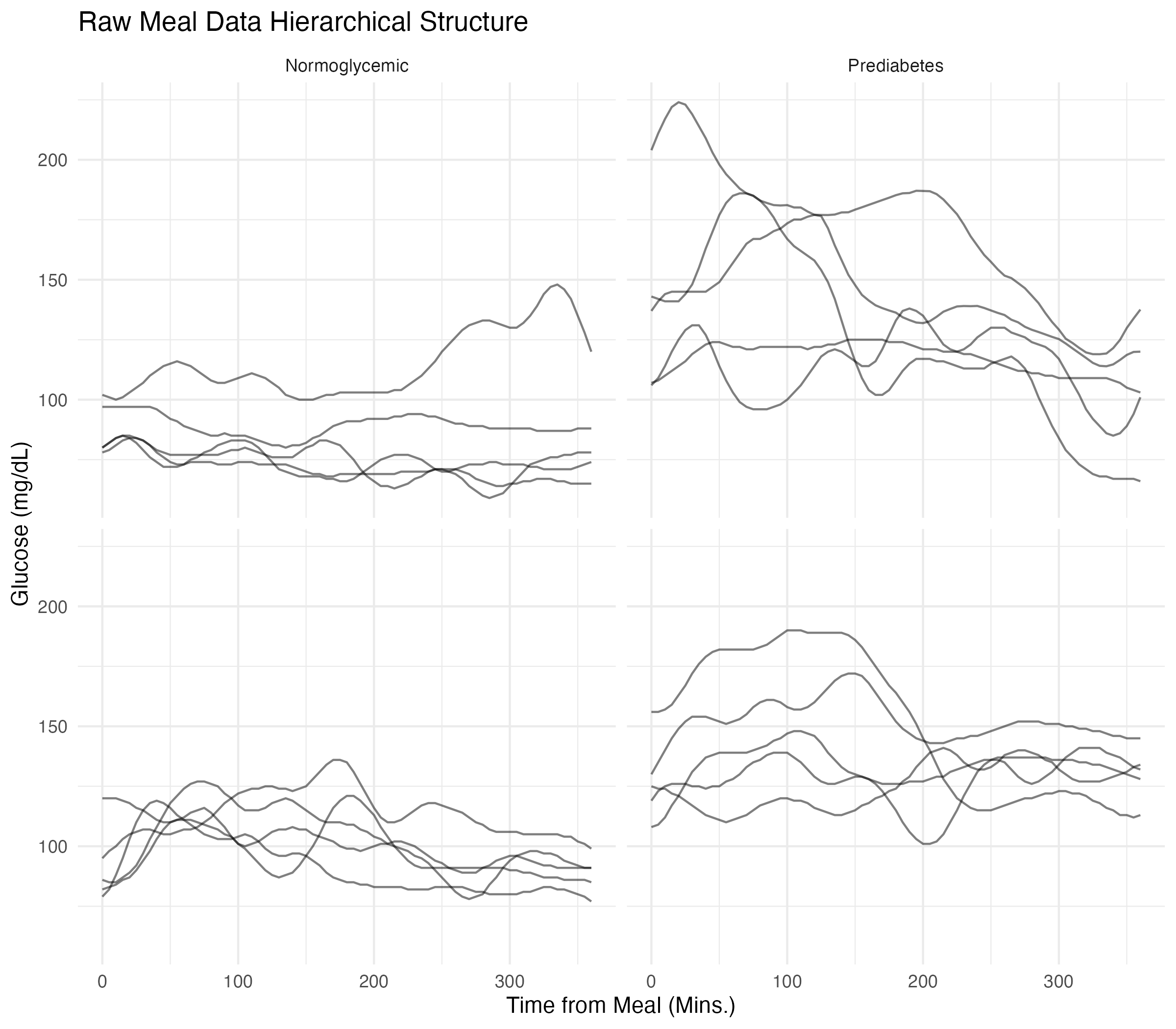}
  \caption{Post-meal raw CGM data over several days in four distinct participants. Each panel corresponds to one participant, and columns indicate glycemic condition - normoglycemic and prediabetes. Groups were assigned according to the American Diabetes Association (ADA) classification, based on glycated hemoglobin (HbA1c) and fasting plasma glucose (FPG) values.}
  \label{fig:RawTrajectories}
\end{figure}

Functional Data Analysis (FDA) \citep{crainiceanu2024functional} is a branch of statistics that provides methods for analyzing functions instead of, or along with, scalar values. FDA is therefore a natural framework to apply to the continuous, smooth postprandial glucose data produced by CGM devices. FDA techniques can mitigate the analytical limitations that arise when CGM data are reduced to a small set of discrete summaries and can enhance our understanding of postprandial glycaemic responses through models that preserve the continuous--time nature of the input data \cite{ramsay_functional_2005, morris_biomechanics_2001, greven_longitudinal_2010}. To gain a deeper understanding of the need to perform functional analysis, Figure \ref{fig:RawTrajectories} illustrates glucose response trajectories after meals over six days in free-living conditions for four individuals -- two normoglycemic and two with prediabetes.  This Figure highlights the substantial inter-individual and inter-day variability observed over the six-hour post-meal period. Our aim is to partition this variability into diet-related effects and residual (unexplained) variation. The two pre--diabetic participants show greater day-to-day fluctuations, whereas the four normoglycaemic individuals display more consistent, subject--specific trends. FDA has already emerged as a promising approach for a variety of CGM-data modeling tasks, from analyzing glucose distributions to characterizing nocturnal glycemic patterns \cite{matabuena_glucodensities_2021, matabuena_hypothesis_2023, matabuena_kernel_2022, cui_investigating_nodate, gaynanova_modeling_2020-1, sergazinov_case_nodate}. However, comprehensive examples detailing the application of multilevel functional modeling to postprandial CGM data are rare, despite the utility of FDA in this context.

This article provides practitioners with a step--by--step guide for analyzing postprandial CGM data, with the goal of fostering broader adoption of these statistical techniques in similar datasets. We demonstrate how FDA methods can be applied directly, using pre-existing software, to quickly estimate modes of variability in glucose trajectories and quantify the association between participant characteristics/meal composition and glycemic response trajectories. 

\begin{figure}[ht!]
  \centering
  \includegraphics[width=10cm]{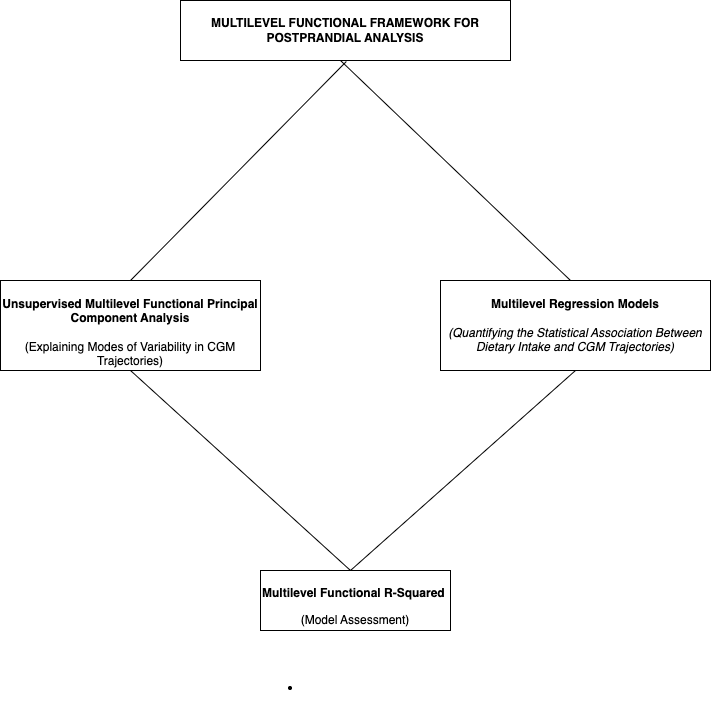}
  \caption{Pipeline Steps for a Multilevel Postprandial Framework.}
  \label{fig:pipe}
\end{figure}

We proceed by providing details of the AEGIS study, explaining the FDA methodologies, and finally demonstrating the results of applying these methods to our hierarchically structured postprandial CGM data. We focus first on descriptive analyses, examining the modes of variability in the postprandial CGM responses using Multilevel Functional Principal Components Analyses (MFPCA). Then, we investigate statistical associations between covariate predictors, including participant characteristics and meal composition, and functional CGM postprandial responses using Function-on-Scalar Regression (FoSR). For both MFPCA and FoSR, it is of interest to understand how much variability the model captures. For MFPCA, explaining a low amount of variability is a good diagnostic indicator that the model structure or truncation are potentially ill-suited to the data. For FoSR, it is important to understand the variability explained by the chosen predictors, both for model selection and to place inferences in wider context. As this is an open area in FDA, we conclude by introducing an extension of marginal and conditional R-square for hierarchical functional models.
Figure~\ref{fig:pipe} summarizes the core steps of the pipeline for analyzing postprandial glucose data using FDA.

 \section{Data: The A Estrada Glycation and Inflammation Study (AEGIS)}

\subsection{Objective and Design}\label{Data:Design}
The A Estrada Glycation and Inflammation Study (AEGIS trial NCT01796184 [Registration Date--2013-02-14]) was a ten-year longitudinal study focused on changes in blood glucose and their connections to inflammation and obesity \cite{gude_glycemic_2017}. The study explored the link between these factors and the potential development of comorbidities such as diabetes mellitus. AEGIS involved a stratified random sample of individuals aged 18 and older, drawn from the Spanish National Health System Registry. The trial further collected CGM for a subsample of the study population, providing detailed glucose profiles at various time points over a period of five years.

At the beginning of the study, a random sample from the general population of 1,516 individuals underwent extensive medical examinations to construct detailed clinical profiles. These included dietary habits, laboratory biomarkers, and responses to questionnaires assessing metabolic capacity, mental well-being, and lifestyles. Table \ref{tab:variables_in_analysis} summarizes all relevant scalar predictors used in our analyses. Distributional summaries are provided for all continuous variables to characterize the study population.

\begin{table}[!ht]
    \centering
    \begin{tabular}{p{4cm} c c}
            \textbf{Variable} & \multicolumn{2}{c}{\textbf{Distribution Summaries}}\\
      \bottomrule
            \multicolumn{1}{c}{} & \text{Normoglycemic }(N=319) & \text{Prediabetes }(N=58)\\
      \bottomrule
            \multicolumn{3}{c}{\textbf{Individual Level\footnotemark[1]}} \\
      \bottomrule 
      Age (yrs) & \makecell[tc]{44.6 (13.7) \\ 44.0 [18.0, 81.0]} & \makecell[tc]{58.7 (12.0) \\ 61.0 [23.0, 84.0]}\\
      \bottomrule 
      Weight (kg) & \makecell[tc]{73.7 (14.3) \\ 72.5 [41.0, 130]} & \makecell[tc]{83.0 (19.6) \\ 79.2 [49.0, 145]}\\
      \bottomrule 
      Gender & \makecell[tc]{Male: 121 (37.9\%) \\ Female: 198 (62.1\%)} & \makecell[tc]{Male: 18 (31.0\%) \\ Female: 40 (69.0\%)} \\
      \bottomrule 
      HbA1c (\%) & \makecell[tc]{5.25 (0.25) \\ 5.30 [3.10, 5.60]} & \makecell[tc]{5.86 (0.20) \\ 5.80 [5.70, 6.40]} \\
      
      \bottomrule 
            \multicolumn{3}{c}{\textbf{Meal Level\footnotemark[2]}}\\
      \bottomrule 
       Carbohydrates (g) & \makecell[tc]{59.9 (40.5) \\ 52.3 [0, 513]} & \makecell[tc]{53.7 (37.5) \\ 45.7 [0, 226]} \\
      \bottomrule 
       fats (g) & \makecell[tc]{30.1 (23.8) \\ 25.4 [0, 237]} & \makecell[tc]{25.7 (22.3) \\ 21.8 [0, 169]}\\
      \bottomrule 
       Proteins (g) & \makecell[tc]{27.5 (17.9) \\ 24.2 [0, 200]} & \makecell[tc]{25.9 (17.0) \\ 23.1 [0.4, 105]} \\
      \bottomrule 
       Fiber (g) & \makecell[tc]{8.8 (6.7) \\ 7.2 [0, 89.3]} & \makecell[tc]{9.1 (7.0) \\ 8.1 [0, 63.1]}\\
      \bottomrule 
       Initial CGM Glucose (mg/dL) & \makecell[tc]{103 (15.3) \\ 101 [52, 237]} & \makecell[tc]{110 (19.4) \\ 107 [56, 196]}\\
      \bottomrule 
    \end{tabular}
    \caption{Description of the variables used collected by AEGIS. Distribution summaries for continuous variates include first Mean (Standard Deviation), followed by the Median [Min., Max.].}
    \footnotetext[1]{All individual-level covariates collected at screening.}
    \footnotetext[2]{All dietary information assessed throughs self-report and dietitian reconstruction.}
    \label{tab:variables_in_analysis}
  \end{table} 

\subsection{CGM and Nutrition Protocols}\label{Data:CGM}

CGM collection was performed for a subset of 581 participants, including 516 individuals without diabetes, in a two-sample design. The other 65 individuals were deemed to have undiagnosed diabetes mellitus and subsequently excluded. Of the remaining individuals, 377 recorded at least one meal which the participant labeled as a "dinner". Our analyses focus on this subset and the meals denoted as "dinners" for the sake of consistency. For each such meal, we extracted the 6 hours of CGM data directly after the reported mealtime to form our postprandial glucose trajectories.

Participants were fitted with Enlite\texttrademark sensors and iPro\texttrademark CGM devices, offering blinded interstitial glucose measurements every 5 minutes for up to seven days. On the seventh day, the sensor was removed, and data excluding the first day's results were downloaded for analysis. Specifics regarding the CGM placement and calibration can be found in \cite{gude_glycemic_2017}.

Throughout the 7-day CGM monitoring period, participants were instructed to maintain their habitual routines while recording every eating occasion in a food diary that spanned the same midnight-to-midnight windows as the sensor. For each meal or snack they noted the clock time, portion size, preparation method (e.g., grilled, fried, baked), ingredients, and any sauces or condiments. At the end of the week a registered dietitian reviewed every diary face-to-face with the participant to clarify ambiguities, fill in omissions, and, when necessary, quantify portions with a validated photographic atlas of household measures; only meals that could be fully decomposed into reliable items were retained. The final diaries were coded in Dietowin® 8.0 (Biologica–Tecnologica Médica, Barcelona, Spain), yielding daily totals for energy (kcal); macronutrients (carbohydrate, sugar, fibre, protein, total fat and its saturated, monounsaturated, and polyunsaturated fractions); cholesterol; and the minerals Ca, Fe, Mg, P, K, and Na.

CGM data were processed conservatively to ensure data quality. The entire first 24 h of wear—during which sensor accuracy was markedly lower (MARD $\approx$ 12 \%)—was discarded. Any subsequent 24-h segment with more than two consecutive hours of signal loss was excluded, thereby removing compression artefacts and extended drop-outs. Each remaining day had to contain at least three capillary finger-stick values for calibration; days failing this requirement were removed without numerical imputation. These rules, together with daily calibrations, limited inter-day drift.

Analyses in the present paper concentrate on evening meals (dinners). Under free-living conditions, dinner is the meal least affected by daytime activity and environmental variability, providing a clearer metabolic signal and greater comparability across individuals. Although the modeling framework can be applied to any meal of the day, our study design and data-quality criteria make dinner the most reliable context for evaluating postprandial glucose dynamics.
 
  In our analysis, we focused on a six-hour period after dinner. Although a six-hour post-prandial window is not standard practice, our functional modelling framework captures glucose responses across the entire time domain; consequently, this  extended window naturally encompasses the conventional two- and three-hour intervals while also capturing longer-term metabolic responses and their statistical associations.

\section{Methods}

\subsection{Notation}\label{Methods:Notation}

The primary data of interest are the hierarchical postprandial CGM curves. We assume that these curves are noisy, discrete observations of some underlying function $Y_{ij}(t)$. We consider index $i \in \{1,\ldots, n\}$ to be the participant - with there being $n$ total. The next index $j \in \{1,\ldots, J_i\}$ indicates the day of observation, where the number of days recorded is $J_i$ for participant $i$. Although standard FDA methods do not require each participant to have the same number of recorded meals -- the AEGIS dataset, for instance, contains heterogeneous counts $J_i$ -- we assume a common value $J_i = J$ across all individuals in this section to simplify the notation. Finally, we consider the domain $t \in [0,360]$, representing the time in minutes since the reported mealtime. In practice, each function is observed only at a finite set of time points, $t \in T_m = \{t_{0}=0,\ldots,t_{m}=360\}$.

\subsection{Multilevel Functional Principal Components Analysis}\label{Methods:MFPCA}

Consider the MFPCA model definition in Equation \ref{eq_MFPCA}, as initially proposed \cite{di_multilevel_2014}. In this model, $\mu(t)$ is the global mean, $\mu(t)+\nu_{j}(t)$ is the mean response for meal/day $j$, $U_{i}(t)$ is the subject-specific deviation from the meal-specific mean function, and $W_{ij}(t)$ is the residual subject- and meal-specific deviation from the subject-specific mean. Here $\mu(t)$ and $\nu_{j}(t)$ are assumed to be fixed functions. While we do not expect there to be any day-specific mean shift $\nu_j(t)$, as there is no consistent pattern in the meals taken by day, we include this term as is standard. We assume that $U_i(t)$ and $W_{ij}(t)$ on the other hand are random functions with expectation $\mathbb{E}[U_i(t)] = \mathbb{E}[W_{ij}(t)] = 0$.

\begin{equation}
Y_{ij}(t)= \mu(t)+\nu_{j}(t)+ U_{i}(t)+W_{ij}(t) \quad \forall i \in \{1,\ldots, n\}, j \in \{1,\ldots, J\}, t\in [0,360]\label{eq_MFPCA}.
\end{equation}

MFPCA proceeds by decomposing $U_i(t)$ and $W_{ij}(t)$ to characterize variability at two hierarchical levels: between and within-participant. Doing so relies upon the functional analog to principal components analysis (PCA). We decompose $U_i(t)$ and $W_{ij}(t)$ in the fashion exhibited by Equation \ref{eq_Decomposition}, where $\phi_{k}(t)$ for $k \leq K$ and $\psi_{h}(t)$ for $h \leq H$ are orthogonal eigenfunctions and $a_{ik}$/$b_{ijh}$ the associated scores or scalar weights.

\begin{equation}
\label{eq_Decomposition}
\begin{split}
U_i(t) & = \sum_{k=1}^{K} a_{ik} \phi_{k}(t) \hspace{0.2cm} \forall i \in \{1,\ldots, n\},\\
W_{ij}(t) & = \sum_{h = 1}^{H} b_{ijh} \psi_{h}(t) \hspace{0.2cm} \forall i \in \{1,\ldots, n\}, j \in \{1,\ldots, J\}.
\end{split}
\end{equation}

The scores are distributed normally with mean zero and variance equal to the eigenvalues associated with each eigenfunction. As in standard PCA, we order the eigenfunctions at each level such that they explain decreasing amounts of the observed variability at their respective levels. These eigenfunctions correspondingly indicate the modes of variability in the observed data at each level. Further, they provide a parsimonious method to represent the observed data: using the corresponding low-dimensional set of scores. 

There is accessible software for estimating the MFPCA model components using data replicates $Y_{ij}(t)$ which are discretely observed over an interval ($t \in T_m$). The \texttt{refund} package provides an efficient method to estimate MFPCA through the function \texttt{mfpca.face} \cite{goldsmith_refund_2020}. This software uses bivariate penalized splines to produce estimates which are robust both to different numbers of observed meals per person and to sparsity in the domain along which the functions are observed.

In this study, we focus on describing the diverse modes of glucose trajectory variability in terms of the eigenfunctions and eigenvalues, facilitating comprehensive understanding of the data structure. We choose the number of eigenvalues $K,H$ at both levels here using our extended R-square measure of variability explained. We estimated this value over a range of $K/H$ values in Supplement \ref{fig:AEGIS:R2_MFPCA_Sense}, where we see a slight elbow at $K = H = 3$, the chosen hyperparameter for our analyses.

\subsection{Function-on-Scalar Regression}\label{Methods:FoSR}

FDA also includes Function on Scalar Regression (FoSR), a method for ascertaining associations between scalar predictors and a functional response. Along with the postprandial glucose $Y_{ij}(t)$, AEGIS collected covariates such as demographics, HbA1c, and meal-level dietary information. We denote these scalar covariate values using $X_{ij,l}$ where index $l\in\{1,2,\ldots,L\}$ refers to a particular covariate. While this notation indicates covariates vary by both person and meal, it is important to note that person-specific features remain fixed over meals.

FoSR presupposes that the outcome function $Y_{ij}(t)$ can be written as the linear combination of the scalar covariates and coefficient functions $\beta_l(t)$. These $\beta_l(t)$ provide an estimate of the association between each predictor and $Y_{ij}(t)$ at time $t$. One can view this as an extension of the basic linear model, where the scalar response and coefficients are generalized to functions. Just as in a traditional linear model, we can extend FoSR to add functional random effects, such that each participant has their own random deviation function. The form of the entire FoSR model, with person-specific random-effects, can be found in Equation \ref{eq_fast}.

\begin{equation}
Y_{ij}(t)= \sum_{l=1}^L X_{ij,l}\beta_l(t) + \alpha_{i}(t) + \epsilon_{ij}(t) \label{eq_fast}
\end{equation}

Within this model, $\alpha_i(t)$ is a random functional effect corresponding to subject $i$ at time $t$. This function plays a similar role to the person-specific deviation from MFPCA, denoted as $U_i(t)$. The function $\epsilon_{ij}(t)$ captures the residual variation that is unexplained by either the fixed or random effects. We assume that the $\alpha_i(\cdot)$ and $\epsilon_{ij}(\cdot)$ processes are zero mean and that $\epsilon_{ij}(\cdot)$ is uncorrelated with all $\alpha_i(\cdot)$, though the $\alpha_i(\cdot)$'s can be correlated among themselves.

Multilevel functional models such as the one in Equation \ref{eq_fast} can be fit efficiently using the R packages \texttt{refund} and \texttt{fastFMM}.
The \texttt{fosr()} family of functions of the former package is best for moderately-sized data. These functions use flexible penalized-spline based modeling to estimate the both the coefficients $\beta_l(t)$ and random effects $\alpha_i(t)$ using reduced rank regression \citep{goldsmith_refund_2020}. This type of adaptive procedure is excellent at recovering the $\alpha_i(t)$, but can be computationally costly. If the data is large though, the \texttt{fui()} function from \texttt{fastFMM} is better suited, as it leverages point-wise models to estimate the coefficients $\beta_l(t)$ \citep{cui_fast_2021}. More recently, scalable methods to estimate both the coefficient functions $\beta_l(t)$ and the random effect functions $\alpha_i(t)$ have been developed \citep{zhou_prediction_2025}. In our empirical analysis, we focus on the fui() methodology to reduce computational cost and obtain results more quickly.

\subsection{$R^2$ for Multilevel Functional Models} \label{Methods:R2}
We extend the traditional notion of $R^{2}$ to multilevel functional models. This extension has two main objectives:  
(i) to assess how accurately the functional models can recreate the response over time, and  
(ii) to guide model selection so that the final model retains sufficient information for the specific application.

R-square, denoted here as $R^2$, is a classical metric in statistical literature used to quantify the variance explained in a response variable by a set of corresponding predictors. For functional data, which is point-wise distributed approximately Gaussian, it is straightforward to extend the traditional formulae for $R^2$ to the functional case by evaluating it repeatedly over the functional domain.

We derive point-wise and global $R^{2}$ metrics for both MFPCA and FoSR. Our extension of the $R^2$ leverages the fact that both FoSR and MFPCA estimate the functional responses $Y_{ij}(t)$ at points $t\in T_m$ over the functional domain. For each $t\in T_m$, $i \in \{1, \ldots, n\}$, and $j\in \{1, \ldots, J\}$, we denote $\widetilde{Y}_{ij}(t)$ and $Y_{ij}(t)$ as the predicted and observed functional trajectories, respectively. At any given time point $t \in T_m$, the point-wise $\widetilde{R}^2(t)$ of the supervised models can be estimated using the standard univariate approach as detailed in Equation \ref{eq_pointwise}. Note that this formula includes averaging over both participants $i$ and meals $j$, forming an aggregate estimate of variance explained over the entire dataset.

\begin{equation}
\label{eq_pointwise}
\begin{split}
\widetilde{R}^2(t) & = 1 - \frac{\sum_{i=1}^{n}\sum_{j=1}^{J} \left(Y_{ij}(t) - \widetilde{Y}_{ij}(t)\right)^2}{\sum_{i=1}^{n}\sum_{j=1}^{J} \left(Y_{ij}(t) - \overline{Y}(t)\right)^2},\\
 \text{where} \ \overline{Y}(t) &= \frac{1}{nJ}\sum_{i=1}^{n}\sum_{j=1}^{J}Y_{ij}(t).
 \end{split}
\end{equation}

We can produce estimates $\widetilde{R}^2(t)$ of variability explained by MFPCA when including just the participant-level components and the whole model. For the participant-level, we form the predicted trajectories $\widetilde{Y}_{ij}(t)$ using the MFPCA estimates at just the first level of the hierarchy. To be specific, we set $\widetilde{Y}_{ij}(t)=\widehat{\mu}(t)+\widehat{\nu}_j(t)+\sum_{k=1}^{K}\widehat{a}_{ik}\widehat{\phi}_k (t)$ for estimated participant-level eigenfunctions $\widehat{\phi}_k(t)$ and scores $\widehat{a}_{ik}$. For the whole model, we correspondingly form the predicted trajectories using all estimates: $\widetilde{Y}_{ij}(t)=\widehat{\mu}(t)+\widehat{\nu}_j(t)+\sum_{k=1}^{K}\widehat{a}_{ik}\widehat{\phi}_k(t) + \sum_{h = 1}^H \widehat{b}_{ijh}\widehat{\psi}_h(t)$. In both cases, we can then apply Equation~\ref{eq_pointwise} to get the estimate $\widetilde{R}^2(t)$ of point-wise $R^2$. 

For FoSR models, we extend the marginal and conditional notions of $R^2$ introduced by \cite{nakagawa_general_2013}. This extension follows directly from the definition of these quantities in the original context of linear mixed effects models. The marginal $\tilde{R}^2$ estimates the variability explained by the fixed factors only, the $\sum_{l = 1}^L X_{ij,l} \beta_l(t)$ component of Equation~\ref{eq_fast}. We can leverage Equation~\ref{eq_pointwise}, setting $\widetilde{Y}_{ij}(t)$ equal to this quantity, to derive the point-wise marginal $\widetilde{R}^2$. The conditional estimate will incorporates both fixed and random factors -- the covariate effects and person-specific deviation $\sum_{l = 1}^L X_{ij,l} \beta_l(t) + \alpha_i(t)$ from Equation~\ref{eq_fast}. The same procedure can be used to derive the point-wise conditional $\widetilde{R}^2$. The respective fitted values used to estimate $R^2$ in each case are detailed in Equation \ref{eq_FoSRFitted}, using the covariates as introduced for Equation~\ref{eq_fast}. 

\begin{equation}
\label{eq_FoSRFitted}
\begin{split}
\widetilde{Y}^{marginal}_{ij}(t) &= \sum_{l=1}^L X_{ij,l}\widehat{\beta}_l(t)\\ 
\widetilde{Y}^{conditional}_{ij}(t)&= \sum_{l=1}^L X_{ij,l}\widehat{\beta}_l(t) + \widehat{\alpha}_{i}(t).
\end{split}
\end{equation}
We define a global estimator of $R^2$ for both FoSR and MFPCA models as the integrated pointwise estimates over the time interval $t\in[0,360]$. Considering the data being observed on a discrete grid, this integral can be approximated using numerical quadrature weights $w(t)$. We leverage simple trapezoidal weights here. The resulting estimator is summarized in Equation \ref{eq_globalR2}. 

\begin{equation}
\label{eq_globalR2}
\widetilde{R}^2 = \frac{1}{360} \int_{0}^{360} \widetilde{R}^2(t) dt \approx \frac{1}{360}\sum_{t \in T_m} w(t)\widetilde{R}^2(t).
\end{equation}
\section{Results}
\subsection{Multilevel Functional Principal Components Analysis} \label{Results:MFPCA}

\begin{figure}[ht!]
  \centering
  \includegraphics[width=10cm]{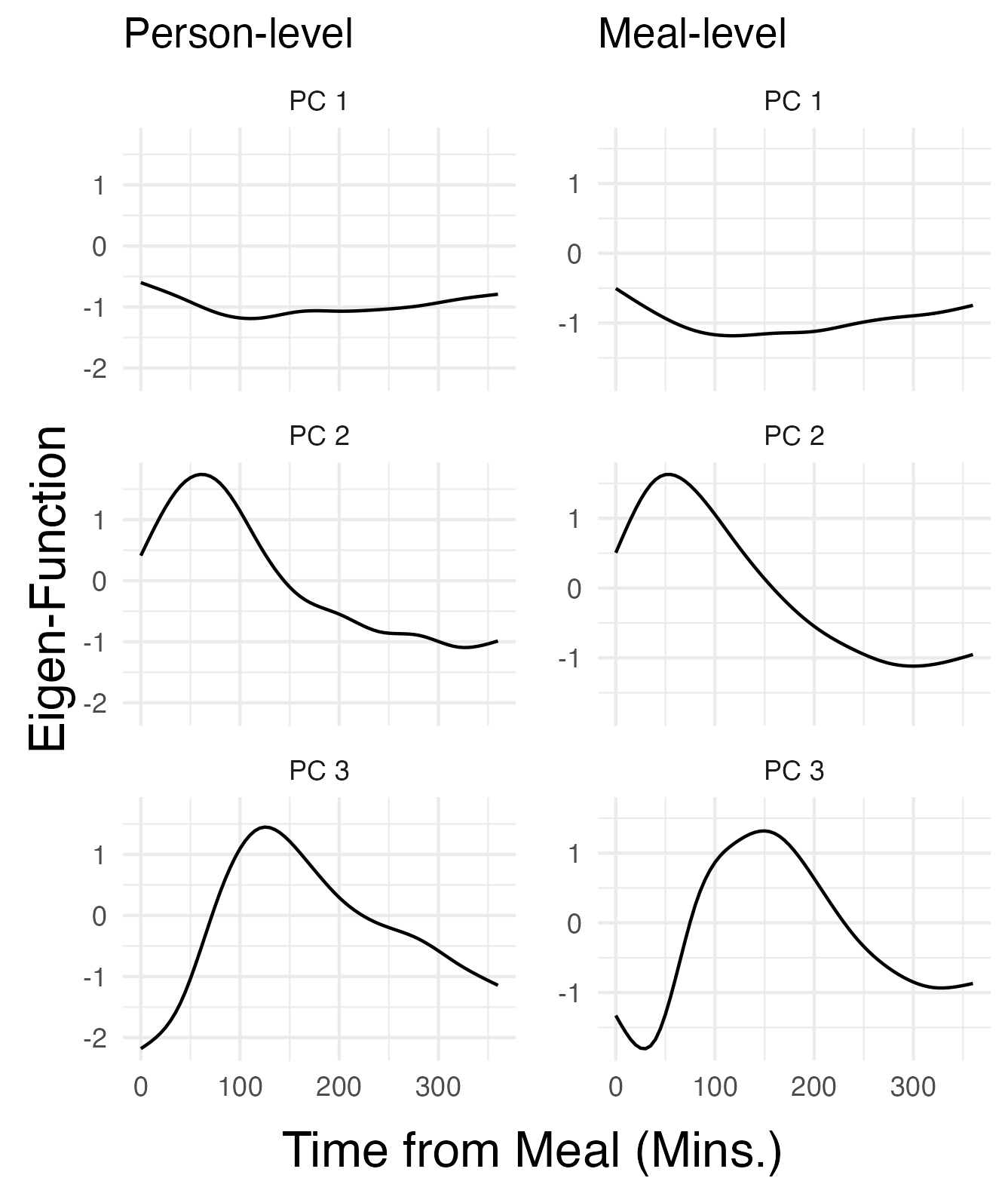}
  \caption{Eigenfunctions of the postprandial CGM responses at the subject and meal levels}
  \label{fig:EigenFuncs}
\end{figure}

We decomposed the postprandial glucose response functions for AEGIS individuals without diabetes into functional principal components. Figure \ref{fig:EigenFuncs} displays the first 3 eigenfunctions at the person- and meal-levels, the 3 orthogonal functions accounting for the most variability in $U_i(\cdot)$ and $W_{ij}(\cdot)$ respectively. The first two eigenfunctions at the individual and meal levels in Figure \ref{fig:EigenFuncs} accounted for a substantial portion of the total variability --- more than 80\%. The first eigenfunctions at both levels suggested an almost time-invariant absolute level, with a relatively shallow concavity peaking at around 100 minutes. The second eigenfunctions contained a more pronounced peak at around 60--80 minutes after the meal. There is also noticeable similarity of eigenfunctions across the two hierarchical levels.

Recall that MFPCA represents all trajectories as the linear combination of eigenfunctions weighted by scores. The first eigenfunction is the one with the greatest variance in the corresponding scores. With this in mind, the first eigenfunction being a close to constant function with small curvature around 100 minutes indicates that the main difference in the postprandial responses, both between and within participants, is the mean glucose level. Further, those with higher mean levels will have slightly higher concavity around 100 minutes. This indicates correlation between higher postprandial response and later peak glucose, which aligns well with physiological understanding of how variation in diet and participant capacity to process insulin can impact glucose processing \cite{brand-miller_glycemic_2009}. The second and third eigenfunctions at both levels can similarly be placed within the context of physiological understanding: (2) how high the postprandial peak (at $\approx$60 mins) goes, with elimination speed of glucose scaling correspondingly to maintain the total processing time; and (3) whether the glucose peak is later, at $\approx$100 minutes rather than $\approx$60. Both of these features can be connected to higher glycemic index of meals or impaired glucose processing in normoglycemic individuals.

\begin{figure}[ht!]
  \centering
  \includegraphics[width=11cm]{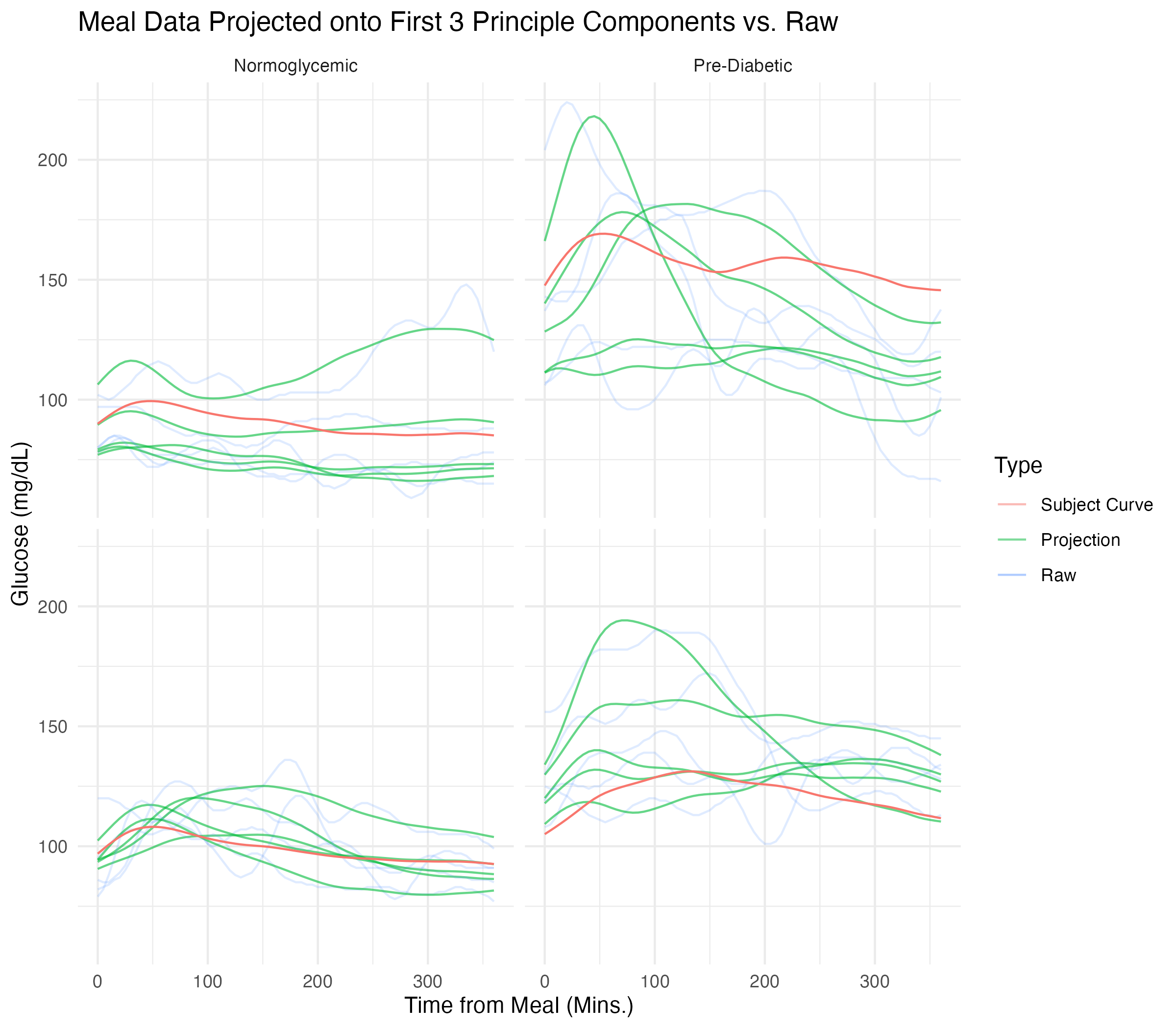}
  \caption{Projections of Raw Data onto Eigen-Functions. \textbf{Raw Trajectories (Transparent Blue):} These trajectories display the original data collected over consecutive days. \textbf{Estimated Participant Trajectories (Red Curve):} Utilizing the scores and eigenfunctions at the individual level, these visualizations illustrate the estimated participant-level trajectories. \textbf{Smoothed Projected Trajectory for Each Meal (Green Curves):} Constructed from the scores and eigenfunctions at both levels, these curves represent the projected meals from each participant.}
  \label{fig:Proj_Onto_EigenFuncs}
\end{figure}

Figure \ref{fig:Proj_Onto_EigenFuncs} shows the CGM raw trajectories for four randomly selected individuals - two with prediabetes and two who are normoglycemic - along with their corresponding person-specific and meal-specific projections according to the MFPCA model. These components were calculated using the relevant estimated components from Equation~\ref{eq_MFPCA}.

We observe pronounced heterogeneity among the projected data at both levels, with variability both between and around the subject-specific trajectories. Eigenvalue analysis indicates that, of the variability explained by this MFPCA model, 33\% comes from the subject level and the remaining 67\% from the meal level, further supporting the hypothesis of important variability existing at both levels. This finding underscores the necessity of using multilevel models to account for CGM data structure.

\subsection{Function-on-Scalar Regression} \label{Results:FoSR}

We applied the Function on Scalar Regression model to examine time-dependent association between covariates of interest and postprandial glucose response. The covariates included those detailed in Table \ref{tab:variables_in_analysis} and initial glucose concentration. This glucose value, measured 5 minutes prior to the recorded meal, was added both to account for relatively high autocorrelation inherent to CGM data due to the nature of glucose dynamics and to introduce some information related to glycemic condition when the meal begins.

\begin{figure}[ht!]
    \centering
    \includegraphics[width=6cm]{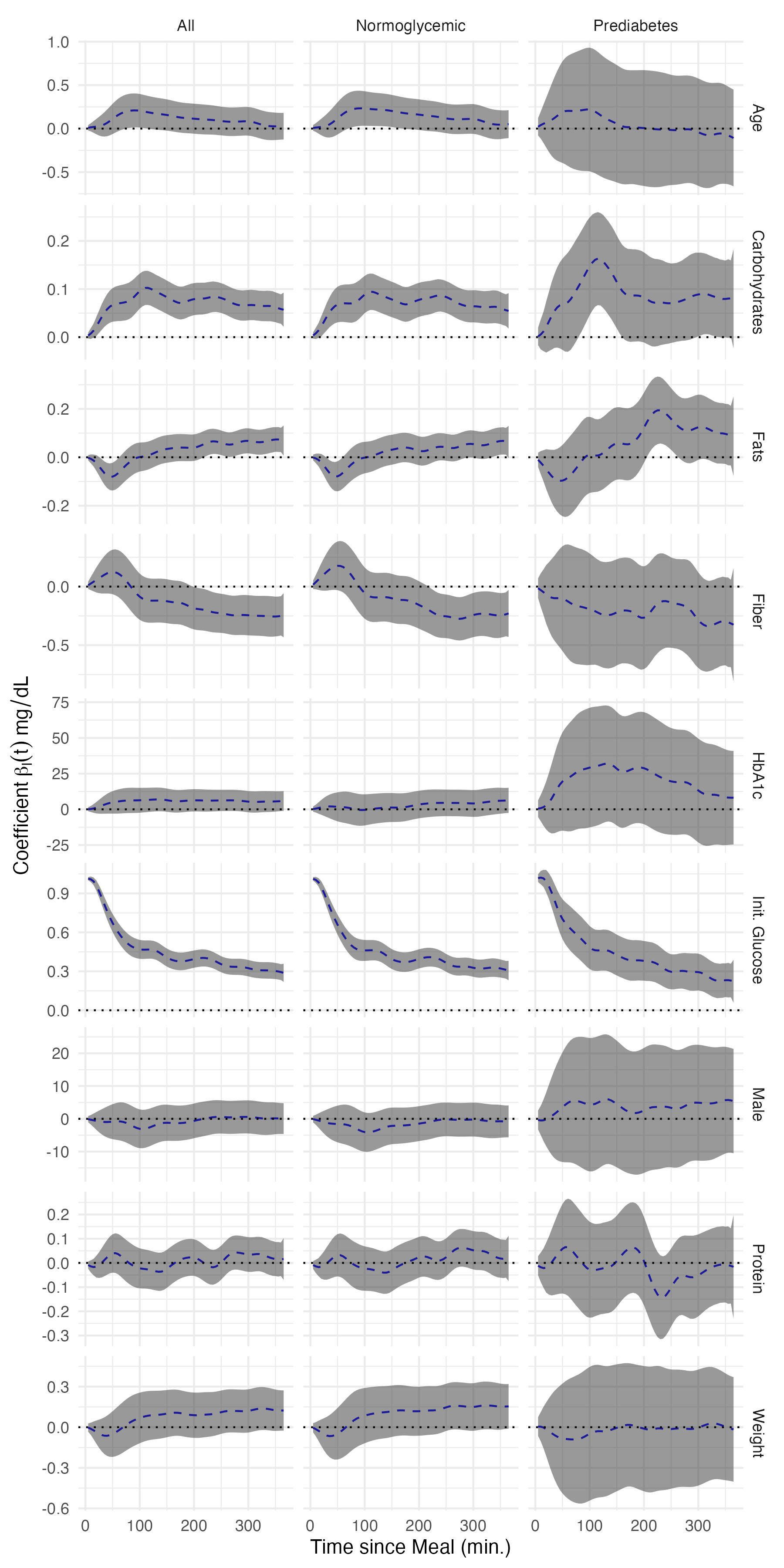}
    \caption{Estimated time‐varying covariate coefficients $\widehat{\beta}_{s}(t)$ from the multilevel function‐on‐scalar regression with postprandial glucose trajectories as the response. Dashed blue curves show the point estimates, grey shading denotes the 95\% confidence bands, and the dotted horizontal line at $0$ marks the null effect.}
    \label{fig:cov_funcs}
\end{figure}

Figure \ref{fig:cov_funcs} displays the fixed effect coefficient functions with associated joint confidence intervals. The first column indicates the coefficient functions for the entire AEGIS population without diabetes (n=377), the second includes just those labeled as normoglycemic (n=319), and the final column contains those individuals with prediabetes (n=58). Each plot contains a dotted line at zero to make it easier to discern where point-wise and joint statistical significance are achieved. As Figure \ref{fig:cov_funcs} makes clear, most covariates achieved point-wise significance over some interval in at least one population, but each had a unique coefficient function shape and subsequent interpretation. We did not standardize the predictors prior to fitting FoSR, as comparison between predictors is not the primary goal and we wish to retain interpretability on the natural scale.

Examining Figure \ref{fig:cov_funcs}, it is first apparent that estimates within the pre-diabetes population were more variable. This was logical given this subset's smaller size and greater heterogeneity in glycemic regulation. With increasing age, there was an increase in postprandial glucose concentrations peaking at 90 minutes. This effect gradually declined until disappearing 5-6 hours after ingestion. No significant differences were observed between men and women. Heightened levels of HbA1c were associated with increase in glucose concentrations along the continuum from normoglycemic to prediabetes, but not within groups. Higher amounts of carbohydrates corresponded to substantial increase in postprandial glycemic response. This effect is magnified in participants with prediabetes, potentially due to lower insulin sensitivity within this group. An opposite effect was observed in meals with higher amounts of fats, where initially (up to 50 minutes post-meal) there is a decrease in glucose concentrations, with a slower and more mild increase in glucose concentration. Again, this effect was greater in individuals with prediabetes than in normoglycemic individuals. Protein intake did not alter post-dinner glucose concentrations; confidence-band analysis showed no statistically significant effects across the entire functional domain. In contrast, fiber consumed about 90 minutes after the post-dinner glucose peak appeared to attenuate the glycemic response, significantly lowering glucose levels in normoglycemic individuals. No significant time-dependent effects were observed in participants with prediabetes, although the pointwise estimates suggested a reduction in glucose values across the entire domain. Initial blood glucose concentration was highly significantly associated with the postprandial response, being particularly influential in the time directly after the meal. While this could be due to high temporal autocorrelation in the CGM data, the observed effect did not decay to zero over the course of the meal window. The higher initial glucose concentrations thus seemed to indicate postprandial glucose concentration in more than an auto-regressive capacity.

\subsection{$R^2$ for Multilevel Functional Models} \label{Results:R2}

We next assessed the variance explained by our hierarchical functional models using a newly introduced notion of functional $R^{2}$, defined from both point-wise and global perspectives. Specifically, we computed $R^{2}$ values for the MFPCA and FoSR models to quantify the proportion of variance explained at each time point $t \in [0,360]$ of the post-prandial period and across the entire functional domain. Defining a functional $R^{2}$ in this context serves two purposes: (i) to determine how much variability is explained by the multilevel functional model--either with or without random effects (ii) to guide model specification, for example by selecting the number of eigenfunctions in MFPCA according to the proportion of variability they capture or by deciding whether to include random effects in the regression setting.

We evaluated $R^{2}$ values for the MFPCA model, integrating over the functional domain with $K = H = 3$ principal components. The corresponding curves are shown in Supplement~\ref{fig:AEGIS:R2_MFPCA}. These curves reveal that the participant--level models exhibit relatively low explanatory power, highlighting substantial intra--participant heterogeneity. In contrast, the full model performs well overall, except for reduced explainability during the first hour after the meal and the final hour of the observation window.

Supplement \ref{fig:AEGIS:MCR2_over_time} was constructed to demonstrate both the marginal and conditional $\widetilde{R}^2(t)$ functions for the FoSR model. For normoglycemic participants, conditional and marginal $\widetilde{R}^2(t)$ values aligned closely in the first 50 minutes, indicating minimal influence of random effects during this period. Later, $\widetilde{R}^2(t)$ values stabilized with random effects contributing to a more than 50\% increase in variability explained. In prediabetic individuals, conditional and marginal $\widetilde{R}^2(t)$ diverged earlier in the postprandial period, potentially a result of the smaller sample of participants with prediabetes, which also has naturally increased heterogeneity in glycemic response. Overall, the models’ predictive capacity is moderate; in particular, after 90 minutes the conditional and marginal versions of $\widetilde{R}^2(t)$ fall below $0.5$.

\section{Discussion}

This paper introduces a FDA framework for studying postprandial CGM responses and demonstrates application of this framework to the AEGIS study. This modeling strategy semi-parametrically estimates time-varying relationships between predictors and the CGM response while still accommodating participant-specific random effects. To the best of our knowledge, it is the only FDA-based framework in the postprandial glucose literature that incorporates random effects. FDA models, augmented with the notion of functional $R^2$ introduced in this paper, provide two key insights in the context of AEGIS: (1) the importance of accounting for participant-specific random effects; and (2) the estimated impacts of diet and participant characteristics on CGM-measured postprandial glucose response trajectories. 

There is a large body of literature modeling glycemic responses to food intake. These models are most often based upon differential equations modeling of the various components of the glycemic system \cite{bergman_origins_2021, urbina_mathematical_2020, maas_physiology-based_2015, shi_modeling_2020, de_gaetano_modeling_2021, eichenlaub_glucose-only_2021, zhang_data_2016, eichenlaub_model_2019, trajanoski_fuzzy_1996, holtschlag_state-space_1998}. There are very few such methods which can comprehensively model the glucose time series data from CGM in isolation of insulin and the other hormones dictating glucose dynamics. Among such models on glucose, the FDA approach is unique in that it is fully data-driven, capable of handling hierarchical structure, and able to estimate time-dependent, semi-parametric associations with scalar predictors (the coefficient functions $\beta_l(t)$). 
Furthermore, the FDA methods we introduce are computationally scalable and can be applied to large medical-cohort studies, including those currently underway in Israel and the United States \cite{shilo_10_2021,noauthor_all_2019}.

Applying FDA to the AEGIS data yields statistically significant, time-varying associations between diet composition and postprandial glucose response. These results were differential by prediabetes status. Importantly, we observed differential glycemic response to increased fat intake between normoglycemic participants and those with prediabetes. Further scientific findings discussed in this section are outlined in Table \ref{tab:findings}. The dietary findings of this study indicate the potential for greater glucose control through combining better understanding of the dietary impact on postprandial glucose and personalized nutritional recommendations.  

\begin{table}[h!]
\centering
\begin{tabular}{>{\raggedright\arraybackslash}p{7cm}>{\raggedright\arraybackslash}p{7cm}}
\textbf{Result} & \textbf{Implication} \\
\bottomrule
There is substantial heterogeneity in mean value and shape of postprandial glucose time series both between and within individuals & Appropriately accounting for the hierarchical structure of the postprandial responses is required for adequate explanation of the observed glucose patterns \\
\bottomrule
The coefficient functions $\beta_l(t)$ for different macro and micro-nutrients are not time-invariant, and they vary in intensity and direction. & Postprandial glucose response is influenced by the composition of macro and micro-nutrients in distinct ways. \\
\bottomrule
The coefficient functions $\beta_l(t)$ differ between normoglycemic and prediabetic individuals. & Metabolic responses to the same diet differ between normoglycemic and prediabetic patients, indicating the importance of glycemic capacity in formulating diet recommendations. \\
\bottomrule
The $R^2$ functional mixed model explainability metrics are not time-invariant, showing a decline over time, and random effects significantly increase the variability explained in the predictions $50$ min after post-meal intake. & Post-meal functional response analysis indicates significant individual heterogeneity, necessitating alternative, perhaps more personalized, model structures. \\
\bottomrule
\end{tabular}
\caption{Summary of Findings}\label{tab:findings}
\end{table}

The FoSR models fit here explain only a moderate proportion of the variability beyond the first hour, even after incorporating the participant-specific random effects. This suggests the presence of latent structure that is not captured by the covariates we collected. We hypothesize that unmeasured factors such as physical activity and details of proximal snacking (timing and composition) contribute to this residual variability. Capturing a larger share of the variance may therefore require adding a more large list of relevant variables or adopting a more flexible random-effects structure. We also emphasize that modeling the postprandial response is intrinsically challenging, largely because of the considerable inter--individual metabolic variability.

Regarding FoSR model assessment and potential over-fitting, we note that the underlying specification remains linear. This simplicity provides an intrinsic safeguard against over-fitting. Linear models are generally robust, though they can be sensitive to mis-specification because of their rigid functional form. The predictive variability observed among participants with pre-diabetes reflects true inter-individual heterogeneity within this subgroup and the fact that we analyzed a relatively small subsample ($n = 58$). As future work, we plan to develop more flexible functional models that can capture general non--linear, time--dependent associations between diet and glucose trajectories--ranging from advanced statistical approaches to machine-learning methods--while carefully balancing the increased risk of over-fitting.

In this paper, we focus on the $fui()$ function because it delivers inference comparable to state-of-the-art functional methods while remaining computationally lightweight, enabling the analysis of thousands of CGM curves within seconds. For smaller sample sizes ($n < 30$), where inference is more sensitive, a one-step multilevel functional model implemented via $fosr()$, or Bayesian approaches that incorporate expert knowledge through informative priors, may be preferable.

A primary strength of this particular application of FDA to the AEGIS study is the generalizability of results. The AEGIS trial includes a random sample of the general Spanish population, and the normoglycemic sub-group is of sufficient size. The results in this group should generalize well to similar such populations.

The FDA techniques described here can be applied to CGM data from a wide range of populations. Extending the methods to cohorts comprising individuals with normoglycemia, pre-diabetes, or non--insulin-treated type 2 diabetes is straightforward. In groups with type 1 diabetes or insulin--treated type 2 diabetes, however, it is essential to incorporate participant-specific treatment regimens, particularly the timing of insulin administration. Once these factors are properly modeled, FDA yields the same depth of characterization and insight that we achieved for the AEGIS cohort.

\section{Conclusion}\label{conclusion}

Our FDA-based approach to modeling postprandial CGM data provides a method of estimating the time-varying associations between dietary/other factors and PPGR, all while accounting for the hierarchical structure of multiply-observed data. We extend the general modeling framework by introducing an extension of the traditional $R^2$ measure to the case of multilevel functional models. The corresponding results indicate the key importance of including person-specific random effects when modeling these data, as well as the room for improvement even after accounting for these effects. With the increasing prevalence of large cohort studies including wearables data from devices such as CGM, actigraphy, and electrocardiogram (ECG), this framework provides a core set of tools for analyzing and characterizing the corresponding multilevel, functional data.

For future work, we propose extending the glucotype concept using multilevel functional models \cite{hall_glucotypes_2018} and developing novel multilevel functional regression models.

\section{List of Abbreviations}

\begin{itemize}
\item CGM: Continuous glucose monitoring
\item AEGIS: A Estrada Glycation and Inflammation Study
\item FDA: Functional Data Analysis
\item MFPCA: Multilevel Functional Principal Components Analysis
\item FoSR: Function on Scalar Regression
\item ECG: electrocardiogram
\end{itemize}

\backmatter

\bmhead{Acknowledgements}

The A Estrada Glycation and Inflammation Study (AEGIS) group would like to acknowledge the efforts of the participants and thank them for participating.

\section*{Declarations}

\subsection*{Ethics approval and consent to participate}

The study procedures adhered to ethical standards, with informed consent obtained from all participants. The study was approved by the Regional Ethics Committee (Comité Ético de Investigación Clínica de Galicia, registration code: 2012/025) and was conducted in accordance with the Helsinki Declaration.

\subsection*{Consent for publication}

Not applicable

\subsection*{Availability of Data and Materials}

While the full AEGIS dataset is not publicly available, a small subset of the data used for these analyses (N = 30), along with the corresponding code can be provided upon request.

\subsection*{Competing Interests}

The authors have no competing interests to declare.

\subsection*{Funding}

This study has been funded by Instituto de Salud Carlos III (ISCIII) through the projects PI11/02219, PI16/01395, PI20/01069 and by the European Union (Programme for Research and Innovation, 2021-2027; Horizon Europe, EIC Pathfinder: 101161509-GLUCOTYPES); and the Network for Research on Chronicity, Primary Care, and Health Promotion, ISCIII, RD24/0005/0010, 20 co-funded by the European Union-NextGenerationEU.

Mr. Sartini was supported by Grant Number T32 HL007024 from the National Heart, Lung, and Blood Institute, National Institutes of Health.

\subsection*{Author contributions}

MM and JS contributed to the conception of analyses, performed data analysis, participated in development of the novel $R^2$ metrics, contributed to result interpretations, and assisted in drafting/editing the manuscript

FG provided insight into the data (including collection practices, etc.) and participated in interpretation of results.

\begin{appendices}

\section{AEGIS $R^2$}\label{Supp:AEGIS:R2}

\begin{figure}[ht!]
    \centering
    \includegraphics[width=8cm]{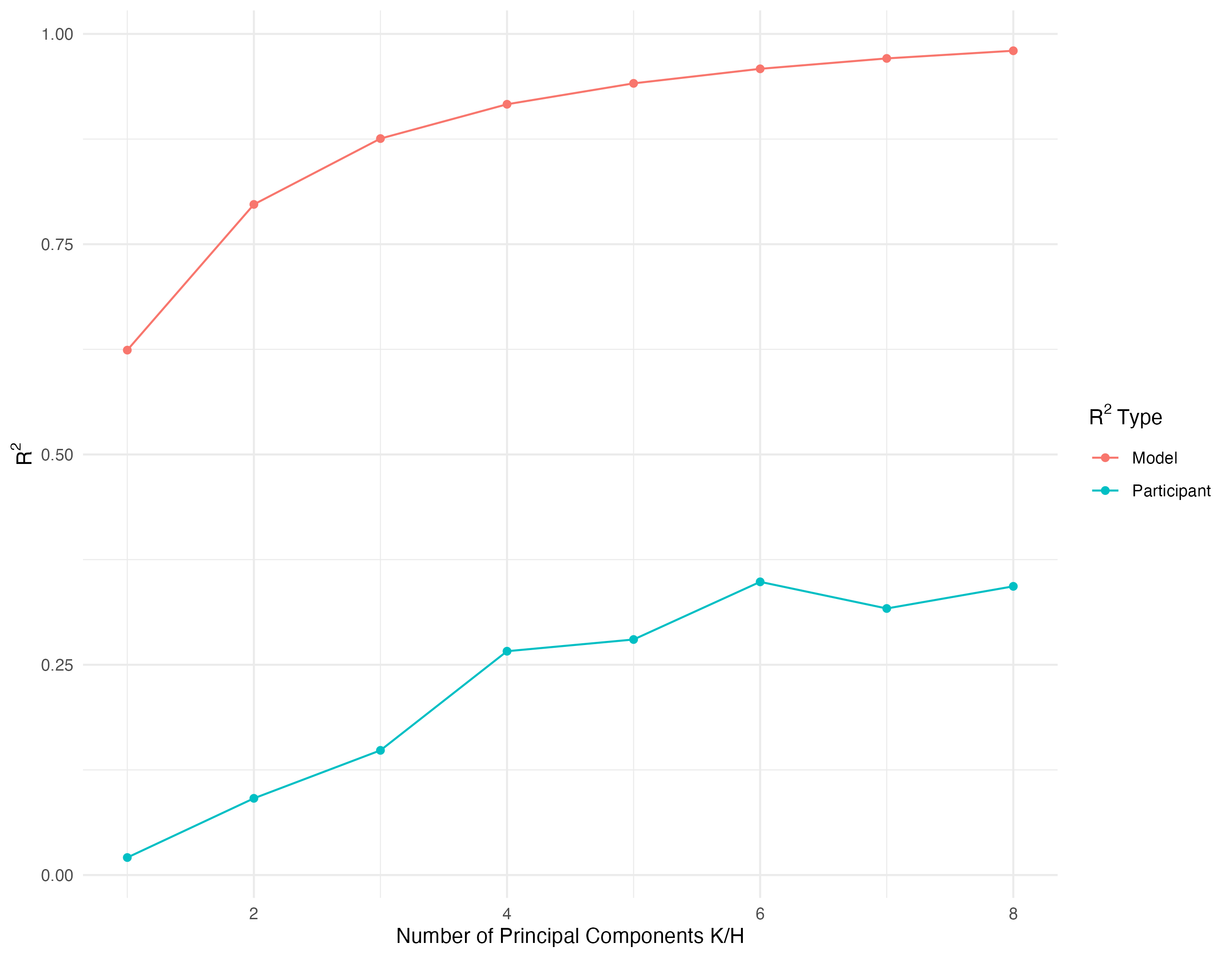}
    \caption{Participant-level and model-level global $R^2$ estimates for MFPCA models by number of eigenfunctions $K,H$ at both levels}
    \label{fig:AEGIS:R2_MFPCA_Sense}
\end{figure}

\begin{figure}[ht!]
    \centering
    \includegraphics[width=8cm]{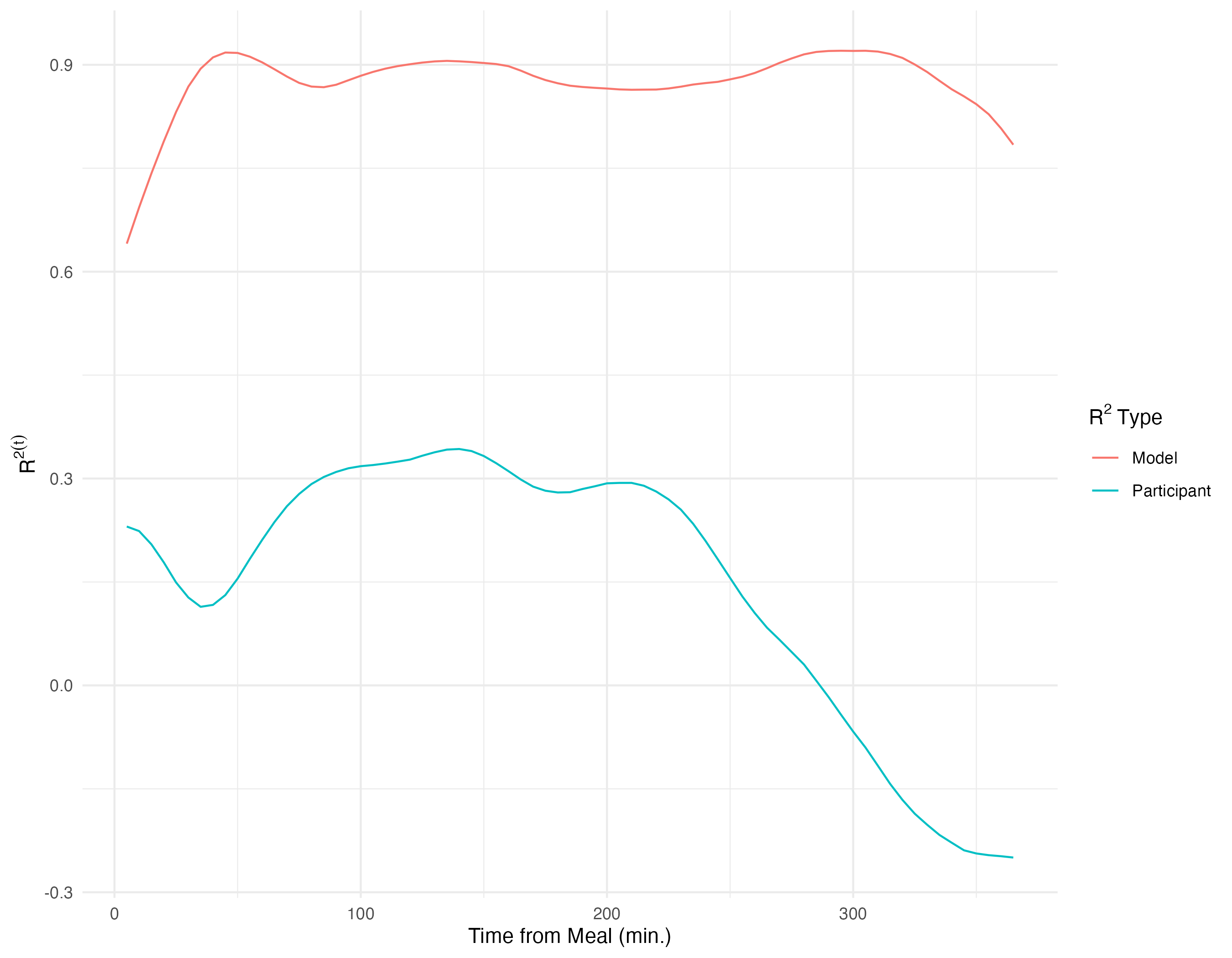}
    \caption{Participant-level and model-level $\widetilde{R}^2$ estimates for the chosen MFPCA model with $K = H = 3$ eigenfunctions at both levels}
    \label{fig:AEGIS:R2_MFPCA}
\end{figure}

\begin{figure}[ht!]
    \centering
    \includegraphics[width=8cm]{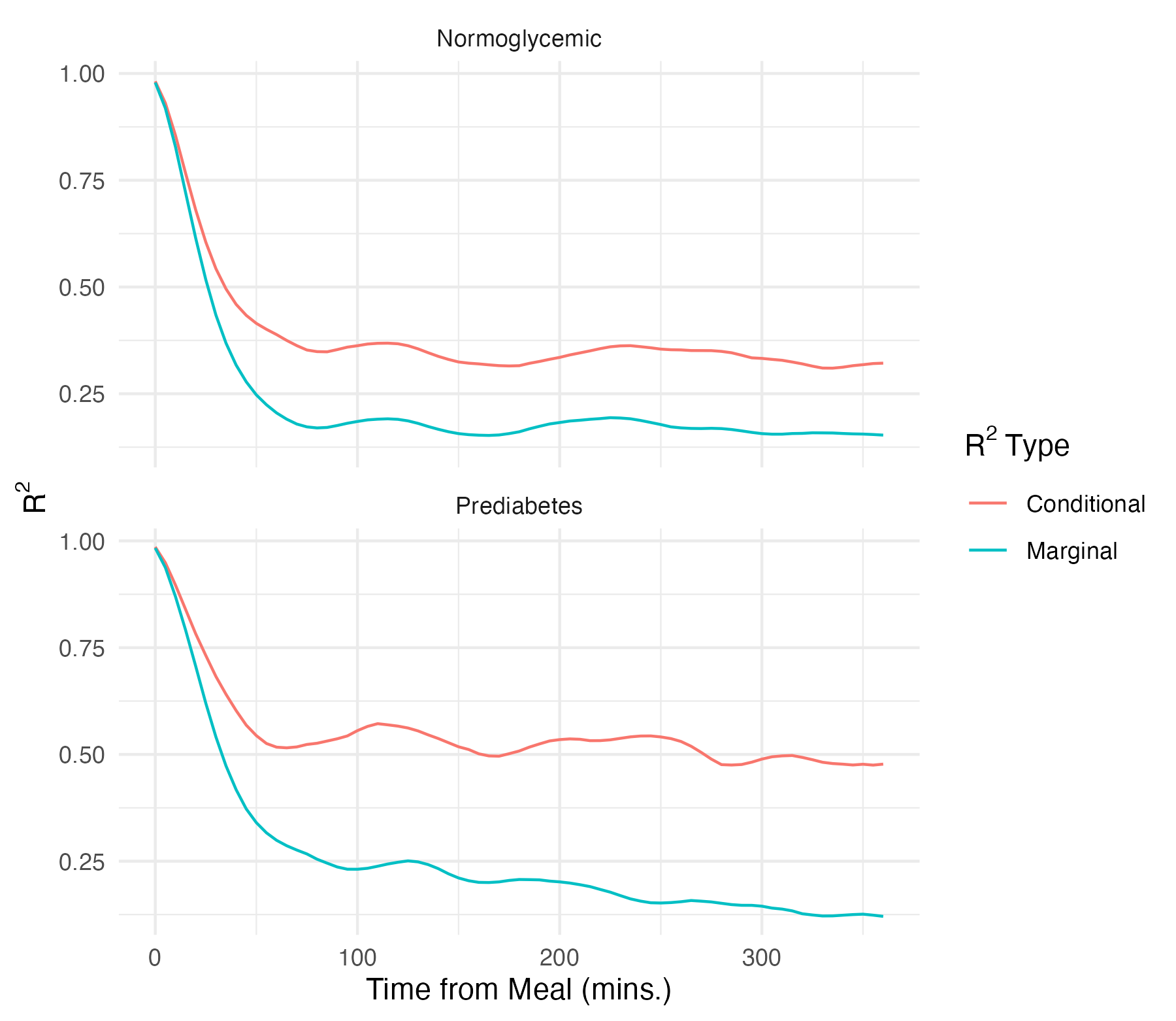}
    \caption{Marginal and Conditional $R^2$ plotted over the course of the postprandial observation window}
    \label{fig:AEGIS:MCR2_over_time}
\end{figure}

\end{appendices}

\newpage

\bibliography{main}
\end{document}